\documentclass[amsmath,nofootinbib,twocolumn,superscriptaddress]{revtex4-1}
\pdfoutput=1
\usepackage[colorlinks]{hyperref}
\usepackage{aas_macros,graphicx,marvosym,subfigure,amssymb,amsmath}
\usepackage{comment}
\usepackage{slashed}

\newcommand{\pd}{\,\partial}

\newcommand{\rs}{r_\star}
\newcommand{\rsh}{r_\sharp}

\newcommand{\SG}[1]{\textcolor{red}{\textsf{[SG: #1]}}}

\begin{document}

\title{Horizon Instability of Extremal Kerr Black Holes: \\ Nonaxisymmetric Modes and Enhanced Growth Rate}

\author{Marc Casals \footnote{}}
\affiliation{Centro Brasileiro de Pesquisas F\'isicas (CBPF), Rio de Janeiro, 
CEP 22290-180, 
Brazil.}
\affiliation{School of Mathematical Sciences and Complex \& Adaptive Systems
Laboratory, University College Dublin, Dublin 4, Ireland.}

\author{Samuel E. Gralla}
\affiliation{Department of Physics, University of Arizona, Tucson, Arizona 85721, USA}

\author{Peter Zimmerman}
\email{peterzimmerman@email.arizona.edu}
\affiliation{Department of Physics, University of Arizona, Tucson, Arizona 85721, USA}

\begin{abstract}
We show that the horizon instability of the extremal Kerr black hole is associated with a singular branch point in the Green function at the superradiant bound frequency.  We study generic initial data supported away from the horizon and find an enhanced growth rate due to nonaxisymmetric modes.  The growth is controlled by the conformal weight $h$ of each mode.  We speculate on connections to near-extremal black holes and holographic duality.
\end{abstract}

\maketitle

\section{Introduction}
As an important and challenging problem, the stability of black holes has attracted considerable interest over the last fifty years.  The main results come in two flavors: mode stability and linear stability.  Mode stability refers to boundedness of perturbations with definite frequency, while linear stability (a stronger result) refers to boundedness of perturbations arising from generic initial data.  We may further differentiate based on the perturbation equations studied, with the main cases of interest being scalar, electromagnetic, and gravitational field perturbations.
  
For the Schwarzschild metric, mode stability was proven in the early seventies \cite{Vish}, followed shortly by linear stability in the scalar case \cite{Kay} and much more recently by linear stability in the gravitational case \cite{Dafermos:2016uzj}.  For the Kerr metric, progress was slower, but the proof of mode stability (for massless fields) \cite{Whiting:1988vc} was eventually followed by linear stability for non-extremal black holes in the massless scalar case \cite{Dafermos:2014cua}.  The orderly progression of results suggests an attitude, widespread among physicists, that for practical purposes mode stability is good enough, with linear stability sure to follow with enough effort from mathematicians.

In 2010 this attitude came under existential threat with Aretakis' discovery of a horizon instability of extremal black holes \cite{Aretakis:2010gd,Aretakis:2011ha,Aretakis:2011hc,  Aretakis:2012ei, Aretakis:2012bm,Lucietti:2012sf, Murata:2012ct} that exists despite their mode stability.  He proved that sufficiently high-order transverse derivatives of axisymmetric massless fields blow up at least polynomially in time along the horizon.  His technique was unlike any used previously, employing a conserved quantity along the horizon that appears as an integration constant in a late-time integration.  This cleverness and originality aside, even the most mathematically-inclined physicist may scratch her head: Is the instability really invisible to a mode analysis?  Is mode analysis really so deficient?

Our head-scratching physicist may find some relief in the results of this paper.  We show how the horizon instability can be recovered in a mode analysis as a singular branch point in the complex-frequency plane of the causal Green function.  We use the approach to fill in details of the axisymmetric case as well as generalize to nonaxisymmetric modes, which turn out to dominate.  The mode-lover can rest easy: a suitably generalized mode analysis recovers the instability and reveals an enhanced growth rate.

We consider a real massless scalar field $\Phi$ on extreme Kerr with initial data supported away from the horizon.  The main result is the growth/decay rate of each angular mode and its transverse derivatives at late times $v$ along the future horizon.  For the $n^{\textrm{th}}$ derivative of the mode labeled by multipole number $\ell$ and azimuthal number $m \neq 0$ (see below \eqref{twenty-seven} for the precise definition), the late time behavior on the future horizon $\mathcal{H}$ is
\begin{align}\label{petermeister}
\left. \Phi^{(n)}_{\ell m} \right|_{\mathcal{H}} & \simeq v^{n-\textrm{Re}[h]}, \quad v \rightarrow \infty. 
\end{align}
 The notation $\simeq$ means asymptotic equality up to a multiplication by a non-zero coefficient $C e^{i f(v)}$ for $v$-independent complex $C$ and real $f(v)$.  (That is, $A \simeq B$ means $A \sim C e^{i f(v)} B$, where $\sim$ is asymptotic equality.) Here $h$ is the so-called \text{conformal weight} that labels representations of the near-horizon enhanced isometry group and plays a key role in the conjectured duality to conformal field theory \cite{Bardeen:1999px,Guica:2008mu,Bredberg:2009pv,Gralla:2015rpa}.  When expressed in terms of the separation of variables normally done in Kerr, it becomes
\begin{align}\label{h}
h=\frac{1}{2} + \sqrt{\frac{1}{4}+K_{\ell m} - 2m^2},
\end{align}
where $K_{\ell m}$ is the spheroidal eigenvalue, defined below.  The appearance of the conformal weight is a hint that deeper understanding may lie in study of the near-horizon conformal symmetries.

We also study axisymmetric modes ($m=0$), revealing the detailed growth/decay rate structure [Eq.~\eqref{truly-truly-truly-outrageous} below].  For generic initial data, all modes will be excited and only the most dominant will survive at late times.  The largest growth rate occurs in the nonaxisymmetric case when the quantity under the square root in \eqref{h} is negative, which occurs for all $\ell$ at sufficiently large $m$ \cite{Yang:2013uba}.  The generic late-time behavior is thus
\begin{align}\label{enhanced}
\left. \Phi^{(n)}\right|_{\mathcal{H}} \simeq v^{n-1/2}, \quad v\rightarrow \infty.
\end{align}
In particular, the value of $\Phi$ falls off like $1/\sqrt{v}$, while the first radial derivative grows like $\sqrt{v}$.  The axisymmetric modes grow only after $\ell+3$ derivatives are taken, and the growth is only $v^{n-(\ell+2)}$.  Thus the actual growth of the instability is faster, and occurs at smaller numbers of derivatives, than would have been predicted based on axisymmetric results.

In the remainder of the paper we derive these results and discuss their significance.

\section{Laplace transform}
The Laplace transform is widely used for stability analysis in engineering and was introduced into the field of black hole perturbation theory in the applied mathematics tour-de-force of Leaver \cite{Leaver:1986gd}.  The transform is defined as
\begin{align}\label{nonnasal}
\tilde{f} = L[f]=\int_0^{\infty} f(v) \ \! e^{-s v} dv,
\end{align}
and the inversion formula is
\begin{align}\label{inverse}
f = L^{-1}[\tilde{f}]= \frac{1}{2\pi i} \int_{c- i \infty}^{c+i \infty} \tilde{f}(s) \ \! e^{s v} ds ,
\end{align}
where $c \in \mathbb{R}$ is taken sufficiently large so that the contour is to the right of all singularities of $\tilde{f}$.  Below we will write $s=-i \omega$ to be consistent with the standard assumed form of a mode expansion, but we emphasize that we use the Laplace, rather than the  Fourier transform.

The Laplace transform is convenient for stability analysis because the late-time behavior of $f(v)$ is determined by the behavior of $\tilde{f}(s)$ near non-analytic points \cite{Doetsch1974}.  This arises as the inversion contour is deformed to encircle the non-analytic points, whose contribution can often be calculated analytically.  The late-time behavior is determined by the point $s_0$ with the largest real part. (We may always use the shift theorem  $L^{-1}[\tilde{f}(s+a)]=e^{-av} f(v)$ to place the point at $s_0=0$.)  Two cases that will arise in this work are
\begin{align}\label{transform1}
s^N \log s & \quad \rightarrow \quad \frac{-(-1)^N N!}{v^{N+1}}, \quad N \in \mathbb{Z}^+. \\
\label{transform2}
s^{-q} & \quad \rightarrow \quad \frac{ v^{q-1}}{\Gamma(q)}, \quad \quad \quad \quad q \in \mathbb{C} \backslash \mathbb{Z}^-,
\end{align}
 where $\mathbb{Z}^\pm$ denotes positive/negative integers including zero. Here the meaning of $A \rightarrow B$ is that if $A$ is the leading non-analytic term in a series expansion of $\tilde{f}$ about $s=0$, then $B$ is the large-$v$ behavior of $f$.  This summarizes the more carefully stated results in Sec.~10.6 of Ref.~\cite{smith1966laplace} and Thm. 37.1 of Ref.~\cite{Doetsch1974}.

\begin{figure}
\includegraphics[width=.5 \textwidth]{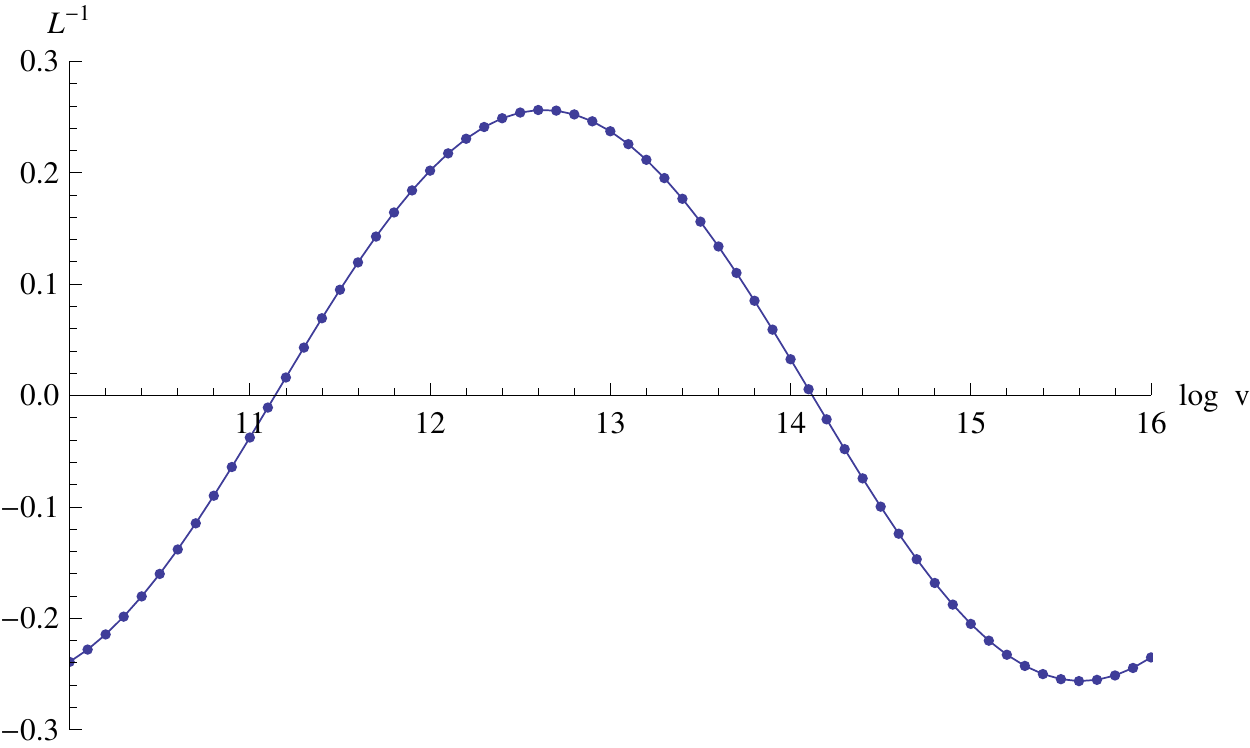}\caption{An example of the inverse Laplace transform \eqref{transform3}.  We plot the real part of $v^{5/2} L^{-1}[s^{-q}/(p+s^{ia})] $ using the parameters $\{p=1.2568474\times 10^{-4}  -4.7826013\times 10^{-5}i    , q = 7/2 - 2.9459584 i,  \alpha =1.8919169 \}$
corresponding to the complex conformal weight associated with the  $\ell=m=2$ mode and the third derivative.  The dots are numerical calculations using the \emph{Mathematica} package \textsf{NumericalLaplaceInversion} based on \cite{VALKO2004629}, while the line is a fit 
to $A \cos(\gamma+\beta\log v)$ (the real part of $v^{5/2}$ times the RHS of \eqref{transform3}) with $A =0.25641175$, $\beta =1.05403570$ and $\gamma = -0.746571461$.  }\label{fig:curvy}
\end{figure}

We will also encounter terms of the form
\begin{align}\label{transform3}
\frac{s^{-q}}{p+s^{i \alpha}} \quad \rightarrow \quad A e^{ i \gamma} \, v^{{\rm Re }[q]-1+i \beta}
\end{align}
where $p\in \mathbb{C}$, $q \in \mathbb{C} \backslash \mathbb{Z}^-$ and $\alpha \in \mathbb{R}$.  We were unable to compute this inverse Laplace transform analytically, and instead established \eqref{transform3} numerically by showing that excellent fits may be obtained for real numbers $A,\gamma, \beta$ for a variety of values of $q$ and $\alpha$---an example is given in Fig.~\ref{fig:curvy}.  The essential point is that the presence of $\alpha \neq 0$ does not modify the overall $v^{\textrm{Re[q]-1}}$ late-time scaling seen in Eq.~\eqref{transform2}.

\section{Green Function}
We consider the extreme Kerr metric in units with $G=c=M=1$, where $M$ is the mass of the black hole.  We use coordinates $\{v,x,\theta,\psi\}$, which relate to the usual Boyer-Lindquist coordinates $\{t,r,\theta,\phi\}$ by
\begin{align}
v = t + \rs, \quad \psi = \phi + \rsh, \quad x = r-1.
\end{align}
where
\begin{align}
\rs= 1+ x-2\left(\tfrac{1}{x} - \ln x\right) ,\quad \rsh= - 1/x.
\end{align}

We will study a massless scalar field,
\begin{equation}\label{wave}
\Box \Phi = 0.
\end{equation}
The solution is given in terms of initial data $\{\Phi_0, n_v \cdot \nabla \Phi_0 \}$   by the Kirchhoff formula
\begin{equation}\label{Kirchoff}
\Phi = \int_{\Sigma} \left ( G \,\,n_v \cdot \nabla \Phi_0 - \Phi_0 \,\, n_v \cdot \nabla G \right),
\end{equation}
where $\Sigma$ is a  smooth hypersurface transverse to $\pd_v$ with future-pointing normal $n_v$.
Here $G$ is the causal Green function, defined to be the solution of
\begin{equation}\label{green}
\Box\, G(X,X') = \delta_4(X,X'),
\end{equation}
that is zero when $X$ is not in the causal future of $X'$, where $\delta_4$ is the covariant delta distribution of spacetime points $X$ and $X'$ \cite{poisson-review}.  This requirement may be imposed by requiring individual frequency modes to be regular on the future horizon and future null infinity, as prescribed by Teukolsky \cite{Teukolsky:1973ha}.\footnote{We are not aware of a mathematical proof of this statement, but it is supported by an enormous body of work making these assumptions and finding causal propagation.}  

We will set $v'=0$ and $\psi'=0$ without loss of generality.  We then mode-expand in the (complete) basis for which the wave equation separates,

\begin{align}\label{bigG}
G & = \frac{1}{2\pi } \sum_{\ell=0}^{\infty} \sum_{m=-\ell}^{\ell}  e^{i m \psi}  \\ & \quad \times \int_{-\infty+ic}^{\infty+ic} S_{\ell m \omega}(\theta)S^*_{\ell m \omega}(\theta')  \tilde{g}_{\ell m\omega}(x,x') e^{-i \omega v} d\omega.\nonumber
\end{align}
The integral is simply the inverse Laplace transform \eqref{inverse} with the notation $s=-i \omega$.  The angular modes are spheroidal harmonics, satisfying 
\begin{align}\label{angular}
  \left[\frac{\partial_\theta ( \sin \theta \, \pd_\theta )}{\sin \theta} + \Big(\hat{K}_{\ell m \omega} - \omega^2 \sin^2 \theta - \frac{m^2}{\sin^2 \theta} \Big)\right]S_{\ell m \omega}= 0.
\end{align}
The spheroidal eigenvalue $\hat{K}_{\ell m \omega}$ is fixed by demanding regularity on both poles. The radial modes $\tilde{g}_{\ell m \omega}$ satisfy the spin-zero Teukolsky equation \cite{Teukolsky:1974yv} with a delta-function source, 
\begin{align}\label{teukolsky}
x^2  &\tilde{g}_{\ell m \omega}''  - i \left(  2\omega x(x+2) +2 i x +  k \right) \tilde{g}_{\ell m \omega}' \\  &+ \left(2 \omega m - 2 i \omega (x+1) - \hat{K}_{\ell m \omega} \right)\tilde{g}_{\ell m \omega} = \delta(x-x'), \nonumber
\end{align}
 where we introduce
\begin{align}\label{k}
k = 4 (\omega - m/2).
\end{align}
Modes with $\omega k<0$ are superradiant (they extract energy from the black hole), so  $k=0$ is called the superradiant bound frequency.  We will see that the instability has its origin in non-analytic behavior at this frequency.  We will denote the angular eigenvalue at $k=0$ by $K_{\ell m}$,
\begin{align}\label{eingenvalue}
K_{\ell m} = \hat{K}_{\ell m \omega}(\omega=m/2).
\end{align}
These may be computed in \textit{Mathematica} by \textsf{SpheroidalEigenvalue[}$\ell, m, i m/2$\textsf{]}. 

The solution of the radial equation \eqref{teukolsky} near $k=0$ requires matched asymptotic expansions \cite{Teukolsky:1974yv,opac-b1091299}, solving separately for $x \gg k$ and $x \ll 1$ and matching in the regime of overlap.  As the approach is by-now standard, we defer the details of the calculation to appendix \ref{app:mae}.  The key result is the formula 
\begin{align}\label{smalldaddy}
\tilde g_{\ell m \omega} & \sim  f(x') \Bigg[\frac{  e^{i \mu/2} (- i k)^{-H +i \mu} }{ \mathcal{S} k (- i k)^{-2 H }   + \mathcal{U} }  \Bigg]e^{i \mu x /2} \nonumber \\ & \quad  \times  \sum_{j=0}^{\infty}\frac{1}{j!} \left(H-  i \mu  \right)_j \left( 1 -  H -i \mu   \right)_j  \left( \frac{x}{ i k} \right)^{j}.
\end{align}
This formula expresses the leading behavior as $k \rightarrow 0$ at fixed $x/k$, given in an asymptotic series in $x/k$ near zero.  We use the Pochhammer notation $(a)_j=\Gamma(a+j)/\Gamma(a)$ and we have introduced
\begin{equation}\label{mu}
  \mu = \frac{k}{2} + m
\end{equation}
and
\begin{equation}
H = \frac12 + \sqrt{\frac14 + \hat{K}_{\ell m \omega} - 2 \mu^2}.
\end{equation}
Notice that $\mu$, $H$, and $\hat{K}_{\ell m\omega}$ reduce to $m$, $h$, and $K_{\ell m}$ (respectively) when $k=0$.  The dependence on the source point $x'\neq 0$ is given by
\begin{align}
f(x') &= e^{- i \mu x'/2} \big[ \mathcal{P}  \,(x')^{H-1}  {}_1F_1(H + i \mu,2 H, i \mu x') \nonumber \\  &+  \,  \mathcal{Q} (x')^{-H} {}_1F_1(1-H+ i \mu,2(1-H), i \mu x') \big], \nonumber
\end{align}
and the coefficients ($\mathcal{P}$, $\mathcal{Q}$, $\mathcal{S}$, $\mathcal{U}$) are given by
{\small
\begin{subequations}\label{coeffs}
\begin{align}
 \mathcal{P} &= - \frac{i\, \Gamma (2 H-1)}{\Gamma (H-i \mu)},   \label{P} \\ 
\mathcal{Q} &= \frac{ \mu (-i \mu)^{-2 H} \Gamma (2 H) \Gamma (2 H-1) \Gamma
   (1-H-i \mu)}{\Gamma (2-2 H) \Gamma (H-i \mu)^2}, \label{Q} \\
\mathcal{S} &= \frac{ (-i \mu)^{1-2 H} \Gamma (2 H) \Gamma (2 H-1)^2 \Gamma (1-H-i
   \mu)}{\Gamma (1-2 H) \Gamma (H-i \mu)^3}, \label{S} \\
\mathcal{U} &= - \frac{ i \pi    \csc (2 \pi  H)}{\Gamma (1-H-i \mu) \Gamma (H-i \mu)}.   \label{U}
\end{align}
\end{subequations}
}

\section{Asymptotics}
The $n^{\textrm{th}}$ derivative of Eq.~\eqref{smalldaddy}, evaluated at $x=0$, contains the leading behavior of $\tilde{g}^{(n)}_{\ell m \omega}|_{\mathcal{H}}$ near $k=0$, which in turn fixes the late-time behavior of the field on the horizon.  By inspection, the value and all derivatives have a branch point at $k=0$, and sufficiently high-order derivatives will also diverge there.  The details depend on the character of the conformal weight $h$.  From Eq.~\eqref{h}, together with calculated values of $K_{\ell m}$ and analytical arguments, we can establish the following properties 
\begin{align}
h \textrm{ is } \begin{cases}=1/2+i b, & |m| \gtrsim .74 \ell \quad \qquad \! \textrm{(case I),} \\ >1, \,\& \notin \mathbb{Z}, & 0 < |m| \lesssim .74 \ell  \quad \textrm{(case II),} \\ = \ell+1, & m=0 \quad  \quad \ \ \qquad \textrm{(case III).}\end{cases}
\end{align}
Here $b$ is a real number.  The last property follows from the fact that $S_{\ell 00}$ are just Legendre polynomials.  The first two properties are established empirically by numerically computing values of $K_{\ell m}$ and the associated $h$.  The transition between case I and II always occurs near $.74 \ell$, and this becomes exact in the large-$\ell$ limit \cite{Yang:2012he}.

\subsection{Nonaxisymmetric Modes}

In the nonaxisymmetric case $m\neq0$, the $n^{\textrm{th}}$ derivative of \eqref{smalldaddy} taken at $x=0$ is given to leading order in $k$ by
\begin{equation}\label{pokemon}
\tilde{g}^{(n)}_{\ell m\omega}\left(k \rightarrow 0\right) \sim C_n \frac{ (-ik)^{h-1-n+ im} }{\mathcal{S}_0 - i \mathcal{U}_0 (-ik)^{2h-1} } ,
\end{equation}
 where
\begin{align}\label{attribute}
C_n & =   i^{2n-1} f_0(x') \, e^{ i m /2}\left(h -  i m  \right)_n \left( 1-h -i m   \right)_n 
\end{align}
and a subscript $0$ indicates evaluation at $k=0$. 

When $h$ is real (case II) the $\mathcal{U}_0$ term in \eqref{pokemon} is subdominant, and the leading small-$k$ behavior becomes
\begin{align}
\tilde{g}^{(n)}_{\ell m\omega}(k \rightarrow 0) \sim \frac{ C_n}{\mathcal{S}_0} (-i k)^{h-1-n+ i m}.
\end{align}
Noting that $s=-i \omega$ and hence $-i k=4s+2 i m$, this is of the form \eqref{transform2} up to a shift $s \rightarrow s-2 i m$.  We can eliminate the shift using the property $L^{-1}[ \tilde{f}(s + a )]  = e^{-a v} f(v)$, which just introduces a phase to the result.  Combining everything together, the late-time behavior of $g_{\ell m}^{(n)}\equiv L^{-1}[\tilde{g}_{\ell m \omega}^{(n)}]$ is
\begin{align}\label{twenty-seven}
g^{(n)}_{\ell m }(v \rightarrow \infty) \sim \frac{ C_n 4^{h+im-n-1}  e^{- i m v /2}}{\mathcal{S}_0 \Gamma(n+1-h-im)}   \,\, v^{n-h-im}.
\end{align}
This translates directly into late-time behavior of field modes according to \eqref{bigG} and \eqref{Kirchoff}.  Specifically, if we decompose relative to spheroidal eigenfunctions evaluated at $k=0$, i.e. $\Phi = \sum_{\ell,m} \Phi_{\ell m}(v,x) e^{i m \psi} S_{\ell m(m/2)}(\theta)$, then the modes $\Phi_{\ell m}$ behave as quoted in Eq.~\eqref{petermeister}.

For complex $h$ (case I), we have $h=1/2+ib$ for real $b$ and hence $2h-1=2ib$ is purely imaginary.  This makes Eq.~\eqref{pokemon} of the form \eqref{transform3} after $s \rightarrow s-2 i m$.   Taking into account the phase arriving from the shift in $s$ as well as $q=1-h+n-i m$ so that $\textrm{Re}[q]=1/2+n$, the late-time behavior is
\begin{align}
g^{(n)}_{\ell m }(v \rightarrow \infty) \sim D_n e^{-imv/2} v^{i \sigma} v^{n-1/2},
\end{align}
where $D_n \in \mathbb{C}$ and $\sigma \in \mathbb{R}$.  The coefficients $D_n$ and $\sigma$ can be fit numerically if desired.  This is the dominant scaling quoted in Eq.~\eqref{enhanced}.

\subsection{Axisymmetric Modes}
In the axisymmetric case $m=0$, Eq.~\eqref{pokemon} is no longer valid, since the coefficient $C_n$ \eqref{attribute} can vanish on account of $h=\ell+1$.  Instead we return to Eq.~\eqref{smalldaddy} using $\mu=2 \omega$, $k=4 \omega$, and $H=\ell+1+O(\omega^2)$.  In the $\ell=0$ case we also need the correction $H=1-(22/3)\omega^2 + O(\omega^4)$ arising from  $\hat{K}_{00\omega}=(2/3)\omega^2+O(\omega^4)$ \cite{Seidel:1988ue}.  We find\footnote{Note that although the ${}_1 F_{1}$ function is singular at non-positive integer values in its second argument, the  limit $\mathcal{Q} \,{}_1 F_1(1-h+2 i \omega, 2(1-h), i \mu x)$ is finite for integer $h$.} 
\begin{align}\label{gintHfull}
&\tilde{g}^{(n)}_{\ell 0 \omega }(\omega \rightarrow 0) \sim  E_n (- i \omega)^{\ell+2 i \omega} \\ &\times \begin{cases} 2 (-1)^{n+\ell} \Gamma (n-\ell) \Gamma (n+\ell+1)(-i \omega )^{1-n},  & n > \ell, \\  
\frac{ \Gamma(\ell + 1 + n)}{\Gamma(\ell + 1 - n)}(-i \omega)^{-n}, & n \leq \ell, \nonumber
\end{cases}
\end{align} 
 where the $x'$ dependent coefficient is given by $E_n =-2^{-2n}11 (11+ 3x')/(103x')$ when $\ell=0$ and $E_n =-2^{-2n-1} \sqrt{\pi}(x')^{-\ell-1}/ \Gamma(\ell+3/2)  $ when $\ell>0$.  The separate case for $\ell=0$ arises because the $\mathcal{U}_0$ term in \eqref{pokemon} is only subdominant when $\ell>0$.  The separate cases for $n > \ell$ and $n \leq \ell$ arise because of the zeros of the Pochhammer function at negative integer values. 

Equation \eqref{gintHfull} gives the small-$\omega$ behavior of the Green function modes.  The corresponding late-time behavior may be computed from Eqs.~\eqref{transform1} and \eqref{transform2}. For example, when $n=0$ we may write $\tilde{g}_{\ell0\omega}(s \rightarrow 0) \sim  A s^{\ell}(1 + B \ \! s \log s)$ for $s$-independent $A$ and $B$ after discarding subleading terms.  The leading non-analytic term is $AB s^{\ell+1} \log s$, and hence by Eq.~\eqref{transform1} with $N=\ell+1$, the late-time behavior is $AB(-1)^{\ell}(\ell+1)!/v^{\ell+2}$.  That is, the falloff of axisymmetric modes of $\Phi$ is $v^{-\ell-2}$.

Going through all the possible cases in this way, we find that the general behavior is
\begin{equation}\label{truly-truly-truly-outrageous}
g^{(n)}_{\ell 0  }(v\rightarrow \infty) \sim F_{ n \ell} \begin{cases} v^{-2}, & n =\ell+1, \\ v^{n-(\ell+2)}, & \textrm{otherwise}, \end{cases}
\end{equation}
where $F_{  n \ell}$ is a constant straightforwardly determined from Eq.~\eqref{gintHfull}.  The growth begins at $\ell+3$ derivatives, as first shown by Aretakis \cite{Aretakis:2012bm}.  Previous studies of charged, non-rotating black holes observed the same growth/falloff rates for a few choices of $\ell$ and $n$  \cite{lucietti2012horizon,bizon2013remark,Zhang2013ADS}, but the complete expression for Kerr appears to be new.  The special case for $n=\ell+1$ means that $v^{-1}$ never appears, and is instead replaced by an additional copy of $v^{-2}$.  For example, for $\ell=2$, successive derivatives (beginning with the value $n=0$) go like $v^{-4},v^{-3},v^{-2},v^{-2},v^0,v^1,v^2,$ etc. 
 A similar `skip' was observed in the weaker estimates proven in the original work \cite{Aretakis:2010gd}.

\section{Discussion}

We now discuss some of the implications of our results and possible future directions. First, we note that the falloff of the field off the horizon is like $1/v$, as first found in \cite{Glampedakis:2001js} and straightforwardly confirmed by the methods of this paper.  This is in contrast to the $1/\sqrt{v}$ falloff we find on the horizon.  This provides a simple way to think about the horizon instability: since the field decays at different rates on and off the horizon, transverse derivatives must grow with time.

Second, our analysis addresses a puzzle about the extremal limit of certain quasi-normal modes (QNMs) in Kerr.  A QNM is defined to be a smooth solution of the Teukolsky equation \eqref{homogeneous-teukolsky} that is regular on both the future horizon and future null infinity.  Previous work \cite{extremeQNM2,extremeQNM3,extremeQNM} has identified a class of QNMs with frequency $\omega = m/2 + O(\epsilon)$, where $\epsilon=\sqrt{1-a^2}$ for dimensionless spin parameter $a$.  These modes become arbitrarily long-lived in the extremal limit $\epsilon \rightarrow 0$, but also arbitrarily hard to excite \cite{berti2006quasinormal,zhang2013quasinormal}, so their collective limiting behavior involves an infinite superposition of overtones \cite{extremeQNM}.  Adding to the puzzle is that there are \textit{no} QNMs at the limiting frequency $\omega=m/2$ in the extreme Kerr metric, since (non-superradiant) purely real modes are forbidden by flux conservation.\footnote{Consider the volume bounded by the future/past horizon $\mathcal{H}^\pm$ and future/past null infinity $\mathcal{I}^\pm$.  Energy conservation implies that the net flux through the boundary is vanishing.  A QNM has vanishing flux through $\mathcal{I}^-$ and $\mathcal{H}^-$.  A QNM with real, non-superradiant frequency ($\omega\, k \geq 0$ and $\omega \neq 0$) has positive flux across $\mathcal{I}^+$ and non-negative flux across $\mathcal{H}^+$, in violation of the theorem that the total flux must vanish.}  What happens to these modes?  The answer seems to be that they pile up at $\omega=m/2$ and become the branch point that we study.

This suggests that phenomena associated with the modes---such as slow decay \cite{Glampedakis:2001js,extremeQNM} or gravitational turbulence \cite{Yang:2014tla}---will similarly limit to phenomena associated with the branch point.  Since the Aretakis instability is confined to the horizon \cite{aretakis2012decay}, however, this must occur in some very subtle manner.  One possibility is that near-extremal Kerr perturbations behave like the Aretakis instability in a region of width $\sim \epsilon$ near the horizon over times of $\sim \epsilon^{-1}$.  If so, then the energy in the field near the horizon will grow like $(\partial \Phi)^2\sim v$ according to \eqref{enhanced}, and could back-react on the metric for sufficiently small $\epsilon$.  For gravitational perturbations this would excite nonlinear couplings and could potentially trigger the turbulent cascade proposed in \cite{Yang:2014tla}.  Further work is required to understand this potential transient instability of near-extremal black holes.

Our results should also be relevant for studying the Kerr/CFT conjecture \cite{Guica:2008mu} that processes near the horizon of an extremal Kerr black hole have a dual CFT description.  Indeed, we have seen that the conformal weight $h$ controls the growth rate of the modes.  It would be desirable to account for the instability within the CFT, perhaps along the lines of bulk-boundary correlator matching done for the non-extremal BTZ black hole \cite{Birmingham:2001pj}.  We hope that our detailed analysis of the analytic structure of the bulk propagator will prove useful in this regard.

Finally, we note that the technique can be used to study other types of perturbations (e.g. electromagnetic and gravitational) as well as other extremal black holes, such as charged black holes, higher or lower dimensional black holes, or black holes in anti-deSitter  spacetime. For example, applying this method to gravitational perturbations of extremal Kerr shows that the $n^{\textrm{th}}$ derivative of the Hartle-Hawking Weyl scalar $\psi_4$ \cite{Teukolsky:1974yv} grows as $v^{n+3/2}$ on the horizon.  In particular, the spacetime curvature diverges.  In a forthcoming paper we will present the complete late-time behavior of general spin fields both on and off the horizon.  

We are grateful to Stefanos Aretakis, Adam Pound, Leo Stein, and Aaron Zimmerman for helpful conversations.  This work was supported in part by NSF grant PHY--1506027 to the University of Arizona.
M.C. acknowledges partial financial support by CNPq (Brazil), process number 308556/2014-3.

\appendix
\section{Matched asymptotic expansions}\label{app:mae}
To arrive at the Green function modes displayed in Eq.~\eqref{smalldaddy}, we apply the method of matched asymptotic expansions to the scalar Teukolsky equation. The need for the method arises because solutions of the  radial equation \emph{at} $k=0$ do not satisfy regularity conditions at $\mathcal{H}$ and future null infinity. Instead, one must expand the radial equation in two regions: a near region where $x \ll 1$ and a far region where $x \gg k$. The smallness of $k$ ensures the existence of an overlap region $k \ll x \ll 1 $ where the solutions can be matched. 

In the text we used ingoing coordinates $v,x,\theta,\psi$, which are natural for studying behavior on the future horizon.  However, for ease of comparison with other references we work here with Boyer-Lindquist coordinates $t,x,\theta,\phi$ (using $x=r-1$ instead of $r$).  For a mode of the form
\begin{align}\label{mode}
\Phi = R(x) S(\theta) e^{i m \phi - i \omega t},
\end{align}
the angular equation is the same \eqref{angular}, while the radial equation becomes
\begin{equation}\label{homogeneous-teukolsky}
    (x^2 R')'+ V R = 0,
\end{equation}
where the potential $V$ is given by 
\begin{equation}\label{potential}
    V = H(1-H) +\frac{k^2}{4 x^2}+\frac{k \mu }{x}+\mu ^2 x+\frac{\mu ^2 x^2}{4}.
\end{equation}

\subsection{Far Zone}

In the far zone $x \gg k$, we may drop the second and third terms in \eqref{potential}, and the radial equation \eqref{homogeneous-teukolsky} becomes 
\begin{equation}\label{eq:farODE}
 (x^{2} R ' )' +
\Big( H(1-H)  + \frac14 (4+x) x \mu^2 \Big) R = 0.
\end{equation}
The two linearly independent solutions may be written in terms of confluent hypergeometric functions as
\begin{align}\label{Rfar}
R_{\rm far} = & P x^{H-1} e^{- i \mu x/2} {}_1F_1(H+i \mu,2H,i \mu x) \\ 
& \,+ Q x^{-H} e^{- i \mu x/2} {}_1F_1(1-H+i \mu ,2(1-H),i \mu x).\nonumber 
\end{align}
At small $x$ (corresponding to the overlap region), the far zone solution takes the form
\begin{equation}\label{Rfaroverlap}
R_{\rm far} \sim P x^{H-1} + Q x^{-H}, \quad {x \rightarrow 0}.
\end{equation} 
At large $x$ (corresponding to asymptotic infinity), the solution has the asymptotic form
\begin{equation}
 R_{\rm far} \sim C_{\infty} e^{i \mu x/2} x^{-1+i \mu} + D_{\infty}e^{-i \mu x/2} x^{-1-i \mu},  \quad  {x \rightarrow \infty} ,
\end{equation}
where $C_{\infty}$ and $D_{\infty}$ are formed by linear combinations of $P$ and $Q$.  The solution with no incoming radiation ($D_{\infty}=0$) is conventionally called the ``up'' solution.  With a convenient overall normalization, the up solution has
\begin{subequations}\label{PQ}
\begin{align}
P &= i(-i k)^{1-H-i\mu}e^{-i \mu/2} \mathcal{P},  \label{Pup} \\
Q &=  i(-i k)^{1-H-i\mu}e^{-i \mu/2} \mathcal{Q},\label{Qup}
\end{align}
\end{subequations}
where $\mathcal{P}$ and $\mathcal{Q}$ are given in \eqref{P} and \eqref{Q}.

\subsection{Near Zone}

In the near zone $ x \ll 1$, we may drop the fourth and fifth terms in \eqref{potential}, and the radial equation \eqref{homogeneous-teukolsky} becomes
\begin{equation}
( x^2 R')' + \Big( H(1-H)  +\frac{k^2}{4 x^2} + \frac{k \mu }{x} \Big)R = 0.
\end{equation}
The solutions are Whittaker functions,
\begin{equation}\label{Rnear}
  R_{\rm near} = a_1  \, W_{i \mu, H - 1/2}( - ik /x) +   a_2  \, M_{i \mu, H - 1/2}( - ik /x).
\end{equation}
At small $x$ (near the event horizon $x=0$), the solution has the asymptotic form
\begin{align}
R_{\rm near} \sim C_\mathcal{H} x^{- i \mu} e^{ \frac{ i k} {2x}} + D_\mathcal{H} x^{i \mu}e^{-\frac{ik}{ 2x }}, \quad x\rightarrow 0,
\end{align}
where $C_{\mathcal{H}}$ and $D_{\mathcal{H}}$ are formed by linear combinations of $a_1$ and $a_2$.  The solution with no incoming radiation ($D_{\mathcal{H}}=0$) is conventionally called the ``in'' solution.\footnote{Equivalently, we may demand regularity of $\Phi$ \eqref{mode} on the future horizon ($x \rightarrow 0$ fixing $v$ and $\psi$).  Note that it is necessary to use $H$ rather than $h$ (i.e., to keep $k$ finite) at this stage in order for the solution to be properly regular on the horizon.} With a convenient overall normalization, the in solution has
\begin{align}\label{a1a2}
a_1=(- ik)^{-i \mu} e^{-i \mu/2}, \quad a_2=0.
\end{align}

\subsection{Green Function}
Our interest is in the causal Green function, whose modes satisfy no incoming radiation from past null infinity and from the past horizon.  This is straightforwardly constructed by using the in solution for $x<x'$ and the up solution for $x>x'$, with the delta-function at $x=x'$ matching the coefficients.  We must also introduce a factor of $e^{i(\omega \rs - m \rsh)}$ to correct for the fact that $\tilde{g}_{\ell m \omega}$ is defined relative to $e^{i m \psi}e^{-i \omega v}$ instead of the $e^{i m \phi}e^{-i \omega t}$ decomposition used here.  For $x<x'$, the Green function modes are then given by
\begin{align}\label{greenie}
\tilde{g}_{\ell m \omega} (x,x') =\bar{R}^{\rm in}(x)R^{\rm up}(x')/\mathcal{W},
\end{align}
where $\mathcal{W}$ is the $x$-independent Wronskian 
\begin{align}\label{W}
\mathcal{W} = x^2 (R^{\rm in} R^{\rm up}{}' - R^{\rm in}{}' R^{\rm up} ),
\end{align}
and $\bar{R}^{\rm in}$ is given by
\begin{align}
\bar{R}^{\rm in} & = e^{i(\omega \rs - m \rsh)} R^{\rm in}. \\
& = e^{-ik/(2x)} x^{i \mu} e^{i \mu(1+x)/2} R^{\rm in} \label{Rbarin}
\end{align}
Note that $\bar{R}^{\rm in}$ is a homogeneous solution to the ingoing-coordinate Teukolsky equation \eqref{teukolsky}.  With our normalization we have $\bar{R}^{\rm in}=1$ on the horizon $x=0$.

\subsection{Wronskian}

Since our interest is in $x=0$ (the horizon) and $x' \neq 0$ (a generic point off the horizon), we need the in solution in the near-zone and the up solution in the far-zone.  These are available from the analyses of the previous two sections.  However, to compute the Wronskian we must have both solutions in the same region.  The most convenient  region is the overlap region.  The in solution in the near-zone is [Eqs.~\eqref{Rnear} and \eqref{a1a2}]
\begin{align}\label{Rinnear}
R^{\rm in}_{\rm  near} = (-i k)^{-i \mu} e^{-i \mu/2} W_{ i \mu, H-1/2}(- i k/x).
\end{align}
The overlap region is the $x \rightarrow \infty$ asymptotics, which are
\begin{align}\label{Rinoverlap}
  R^{\rm in}_{\rm overlap} = A x^{H-1}+Bx^{-H}
\end{align}
with
\begin{subequations}\label{AB}
\begin{align}
\label{poodle}
A &= (-i k)^{1-H-i\mu} e^{-i\mu/2} \frac{\Gamma (2 H-1)}{\Gamma (H-i \mu)} = P,\\
 B &= (-i k)^{H-i\mu} e^{-i\mu/2} \frac{ \Gamma (1-2 H)}{\Gamma (1-H-i \mu) } .
\end{align}
\end{subequations}
The up solution in the overlap region is given by \eqref{Rfaroverlap} with \eqref{PQ},
\begin{align}\label{Rupoverlap}
R^{\rm up}_{\rm overlap} = P x^{H-1}+Q x^{-H}.
\end{align}
Computing the Wronksian \eqref{W} from \eqref{Rupoverlap} and \eqref{Rinoverlap} gives
\begin{align}
\mathcal{W} & = \left(A Q - B P \right)(1-2H) \nonumber \\ & =  k(-i k)^{-2i\mu}e^{-i \mu} \left( \mathcal{U} + k (-ik)^{- 2H} \mathcal{S} \right). \label{Wronky}
\end{align}
where $\mathcal{U}$ and $\mathcal{S}$ are given in \eqref{U} and \eqref{S}.  This completes the calculation of the ingredients needed to assemble the Green function \eqref{greenie} for $x$ in the near-zone and $x'$ in the far-zone.  It is straightforward to determine the Green function elsewhere by matching in the overlap region.  For example, with our normalization choices the up solution in the near-zone is given by  
$R^{\rm up}_{\rm near}=R^{\rm in}_{\rm near}+(Q-B)(-ik)^{-H} M_{i \mu,H-1/2}(-ik/x)$. 

\subsection{Asymptotic series near the horizon}
Finally we wish to express the Green function in an asymptotic series for $x \rightarrow 0$.  This follows from the asymptotics of the Whittaker W function \cite{NIST:DLMF}, 
\begin{align}\label{Whitty}
W_{i \mu, H-1/2}& (- i k/x)  \sim e^{ik /(2x)} ( - i k/x)^{ i \mu} \\
&\sum_{j=0}^{\infty}\frac{ \left(H-  i \mu  \right)_j \left( 1 -  H -i \mu   \right)_j }{j!} \left( \frac{x}{ i k} \right)^{j}.  \nonumber
\end{align} 
In particular, from \eqref{Rbarin} and \eqref{Rinnear} we have 
\begin{equation}\label{eq:rinbar}
\bar{R}^{\rm in} \sim e^{i \mu x/2} \sum_{j=0}^{\infty}\frac{ \left(H-  i \mu  \right)_j \left( 1 -  H -i \mu   \right)_j }{j!} \left( \frac{x}{ i k} \right)^{j},
\end{equation}
so that $\bar{R}^{\rm in}=1$ on the horizon.  Eq.~\eqref{smalldaddy} of the main text now follows by plugging Eqs.~\eqref{eq:rinbar}, \eqref{Rfar} for $R^{\rm up}$, and \eqref{Wronky} in to the expression \eqref{greenie} for the Green function.

\bibliographystyle{apsrev4-1}
\bibliography{hi}

\end{document}